\documentclass[aps,nofootinbib,twocolumn,groupedaddress]{revtex4}
\usepackage{graphicx}

\begin{document}

\title{Critical reflexivity in financial markets: a Hawkes process analysis}

\author{Stephen J. Hardiman} \email{stephen.hardiman@cfm.fr}
\author{Nicolas Bercot}
\author{Jean-Philippe Bouchaud}
\affiliation{
 Capital Fund Management, 23 rue de l'Universit\'e, 75007 Paris, France\\
}

\date{\today}

\begin{abstract}
We model the arrival of mid-price changes in the E-Mini S\&P futures contract as a self-exciting Hawkes process. Using several estimation methods, we find 
that the Hawkes kernel is power-law with a decay exponent close to $-1.15$ at short times, less than $\approx 10^3$ seconds, and crosses over to a second power-law regime with a larger decay exponent
$\approx -1.45$ for longer times scales in the range $[10^3, 10^6]$ seconds. More importantly, we find that the Hawkes kernel integrates to unity independently of the analysed period, from 1998 to 2011. This suggests that markets are and have always been close to criticality, challenging a recent study which indicates that reflexivity (endogeneity) has increased in recent years as a result of increased automation of trading. However, we note that the scale over which market events are correlated has decreased steadily over time with the emergence of higher frequency trading.
\end{abstract}

\maketitle

\section{Introduction}
\label{sec:intro}

The ``excess volatility puzzle'' is the name coined to describe Shiller's \cite{shiller} and LeRoy and Porter's \cite{leroy} observation that {\it stock prices
move too much to be justified by changes in subsequent dividends}, in contradiction with the Efficient Market Hypothesis. Although hotly debated ever since among 
theorists (see \cite{leroy2} for a short review), many empirical studies have confirmed that a large fraction of the volatility of asset prices cannot be explained by 
changes in their fundamental value. For example, most large price jumps seem to be endogenously generated rather than related to the arrival of news in the market place \cite{cutler,fair,joulin}.
Although excess volatility, crises and crashes seem to be as old as financial markets themselves \cite{mckay,reinhart}, the debate has recently been stoked by the rise of ``High Frequency 
Trading'' which is blamed for having made markets more volatile and prone to large magnitude events such as flash-crashes without the need for exogenous triggers. Another closely related 
debate concerns the financial transaction tax as a tool to curb the excessive volatility of markets. (Most academic studies however suggest that such a tax would in fact {\it increase} volatility \cite{taxvol}.)

Keynes \cite{keynes}, Soros \cite{soros} and others (see e.g. \cite{sornette,jpb}) have argued that a substantial part of market activity is of endogenous, ``reflexive'' origin. 
However, real progress has been hobbled by the lack of statistical tools that would enable one to define and measure precisely the degree of reflexivity in financial markets. Recently, Filimonov and Sornette (FS) \cite{filimonov} have applied the so-called self-exciting Hawkes process formalism to market events, with the aim of quantifying precisely the level of endogeneity and exogeneity in the market. Hawkes processes originally found applications in modelling the clustered occurrences of earthquakes. Intuitively, earthquakes relax accumulated stresses between tectonic plates -- but by doing so other regions are brought closer to their yield stress, leading to an inter-twined pattern of foreshocks and aftershocks. Hawkes processes describe the arrival of these seismic events as a Poisson point process but with a history dependent intensity which increases when an event has just occurred. The same picture lends itself naturally to the arrival of events in the market place. Not surprisingly, the Hawkes formalism has generated a flurry of activity as a means of modelling volatility and market dynamics \cite{hawkesvolatility,hawkesmicrostructure,bacry,bormetti}.

The results of Filimonov and Sornette are twofold. First, their Hawkes process analysis over the period 1998-2010 on the E-mini S\&P futures contract suggests that the degree of reflexivity has 
increased steadily in the last decade, an effect they attribute to the increased deployment of high-frequency and algorithmic trading. Second, when studied over much shorter time intervals (10 minutes), the Hawkes process analysis is found to be able to detect precursors of the flash-crash that happened on May 6th, 2010. Devising such early warning tools is evidently beneficial (in particular for market regulators). The work of FS is, in this respect, in a similar vein to that of Easley, O'Hara and Lopez de Prado who developed their ``VPIN'' indicator \cite{vpin}. 

The aim of the present paper is to revisit the first claim of FS that concerns the secular increase of reflexivity in financial markets. (Although for the moment we reserve judgement on the second claim, which is of course of high interest as well). We find, in agreement with the recent work of Bacry et al. \cite{bacry}, that the ``influence kernel'' of the Hawkes process (see below for a precise definition) decays as a power-law in time, but with two different regimes and a crossover time of around five minutes. When we calibrate the power-law kernel to mid-price changes in the S\&P futures contract in much the same way as FS, but on longer time periods, we find that the total degree of reflexivity has been {\it extremely stable over time} and in fact equal to the critical value beyond with the Hawkes process becomes ill-defined. This is the main explanation for the discrepancy between our own analysis and that of FS, who calibrate an exponentially decaying kernel on 30-minute windows. In doing so, they only pick up short-term reflexivity, which we agree, has increased in recent years.

We consider our result very plausible as they fit well with results previously reported in the literature. For example, as we show in the next section, the power-law decay of the influence kernel is related to long-range memory of the event rate, which for mid-price changes, should be strongly correlated with volatility. The fact that measures of market activity (such as trading volume and volatility) exhibit long-range dependence is one of the best documented stylized facts of financial markets. Furthermore, that markets operate in the vicinity of a critical point has been repeatedly emphasized in the econophysics literature, see e.g. \cite{minoritygame,subtle,joulin,marsili,jpb,toth,felix,parisi}. Within a Hawkes process description, criticality is related to both long-range memory and the prevalence of endogeneity over exogeneity in a precise manner. Finally, many studies conclude that the statistics of financial markets have been remarkably stable over time and assets (see e.g. \cite{zumbach,jpb} and refs. therein) -- a result that would have been hard to reconcile with the strong increase of reflexivity reported by FS.

\section{Hawkes processes: definition and some useful results}
\label{sec:Hawkes}

\subsection{The general case}

A one-dimensional Hawkes process is a non-homogeneous Poisson process $N(t)$ with an intensity $\lambda(t) = E\left[\frac{dN(t)}{dt}\right]$ described by a constant base intensity $\mu$ and a self-exciting term that is a weighted sum over previous events:
\begin{equation}
\lambda(t) = \mu + \int_{-\infty}^t \phi(t-s) dN(s) \label{eq:hawkes}
\end{equation}
The ``influence kernel'' $\phi(\tau) \geq 0$ describes the effect on the instantaneous event rate $\lambda(t)$ of a past event that took place at time $t-\tau$. Assuming stationarity, and taking the expectation of both sides of Eq. (\ref{eq:hawkes}) allows one to compute the {\it average} intensity $\Lambda$, which is the solution of:
\begin{equation}
\Lambda = \mu + \Lambda \int_0^\infty \phi(\tau) d\tau.
\end{equation}
Defining $n \equiv \int_0^\infty \phi(\tau) d\tau$, the above equation only has a meaningful solution provided $n < 1$, in which case the process is stationary with an average intensity $\Lambda = \mu/(1-n) \geq \mu$. When $n = 0$, the Hawkes process is a homogeneous Poisson process with intensity $\mu$.  For $n > 1$, the intensity of the process defined by Eq. (\ref{eq:hawkes}) explodes in finite time. The critical case $n = 1$ is special and will be discussed below. Hawkes processes can be mapped into the well known branching processes
\cite{harris}: exogenous ``mother'' events occur with intensity $\mu$, and give rise to endogenous ``child'' events, which themselves give birth to more child events. The quantity $n$ is the branching
ratio of the process, i.e. the expected number of additional events that any single event triggers. As is well known, branching processes become critical when $n$ reaches unity, and give rise to infinite families with strictly positive
probability when $n > 1$. 

The functional form of $\phi(\tau)$ further defines the type of Hawkes process. A popular form, chosen by FS in their analysis of market data, is the exponential kernel:
\begin{equation}\label{hawkes_exp}
\phi(\tau) = \alpha e^{-\beta \tau} \longrightarrow n = \frac{\alpha}{\beta}.
\end{equation}
Note that $\beta^{-1}$ defines the memory time of the Hawkes process, beyond which past events have a negligible influence on the instantaneous event rate. However, other forms are possible too. 
For example, a power-law kernel with a short-time cutoff $\tau_0$:
\begin{equation}\label{hawkes_power}
\phi(\tau) = \Theta(\tau - \tau_0) \varphi_0  \frac{\tau_0^\epsilon}{\tau^{1+\epsilon}}, \quad \epsilon > 0, 
\end{equation}
where $\Theta(.)$ is the Heaviside function. In this case, the branching ratio is $n = \varphi_0/\epsilon$. 

A quantity that can easily be measured empirically is the auto-covariance of the process, defined as:
\begin{equation}\label{correlation}
\nu(\tau)= E\left[\frac{dN(t)}{dt}\cdot\frac{dN(t+\tau)}{dt}\right] - E\left[\frac{dN(t)}{dt}\right]^2
\end{equation}

It can be shown that the Fourier transforms of $\nu$ and $\phi$ are related through the following equation \cite{bacry,bremaud,hawkes}
\begin{equation}
\hat{\nu}(\omega) = \frac{\Lambda}{\left|1 - \hat{\phi}(\omega)\right|^2}. \label{eq:covkernelrelation}
\end{equation}
Letting $\omega = 0$ above leads to:
\begin{equation}
\int_{-\infty}^\infty \nu(\tau) d\tau =  \frac{\Lambda}{\left(1 - n\right)^2}
\end{equation}

As long as $n < 1$, the average event rate $\Lambda$ is finite and the autocovariance must integrate to a finite constant. 
All stationary processes with short-range (integrable) correlations therefore satisfy the condition $n < 1$ \cite{bremaud}. 
For $n > 1$, on the other hand, the process is non-stationary with infinite mean, and the autocovariance of the process is trivially non-integrable
(because the variance is infinite as well). But a Hawkes process that is stationary, with a finite mean ($\Lambda$) and variance but with 
long-range dependence of the event rate, i.e. with a diverging integral of the autocovariance \cite{daley},{\it must necessarily be critical}! 
The mathematical existence of such a stationary critical ($n=1$) Hawkes process was proven by Br\'emaud and Massouli\'e \cite{bremaud}, 
and it is this process that we find fits well to our empirical data.

\subsection{The critical case} \label{sec:critical}

Let us study this critical case in more detail. Assuming that the Hawkes kernel is a power-law as above, with an exponent $0 < \epsilon < 1$, one finds that the small $\omega \tau_0$ 
expansion of $\hat{\phi}(\omega)$ reads:
\begin{equation}
\hat{\phi}(\omega) = n - A (\tau_0 \omega)^\epsilon
\end{equation}
where $A$ is a complex number, related to $\varphi_0$. We have factored out $\tau_0$, to make A a dimensionless constant. (Note that when $\epsilon > 1$, the first term of the expansion of $\hat{\phi}$ becomes regular, i.e. $\propto \omega$.)
Similarly, for an auto-covariance $\nu(\tau)$ which decays asymptotically as a power-law $\propto \tau^{-\alpha}$ for some $0 < \alpha < 1$, one finds:
\begin{equation}
\hat{\nu}(\omega) = B (\tau_0 \omega)^{\alpha - 1}
\end{equation}
with some real constant $B$. Note that $\alpha < 1$ corresponds to long-range memory, and $\hat{\nu}(\omega \to 0) \to \infty$, necessarily leading to $n=1$, as alluded to above. 
Inserting these expressions into Eq. (\ref{eq:covkernelrelation}), we obtain, for $\omega \to 0$,
\begin{equation}
B (\tau_0 \omega)^{\alpha - 1} \approx  \Lambda |A|^{-2} (\tau_0 \omega)^{-2\epsilon},
\end{equation}
leading to:
\begin{equation}
\alpha = 1 - 2\epsilon, \qquad B = \Lambda |A|^{-2}
\end{equation}
The above relation implies that $\epsilon$ must in fact lie in the restricted interval $]0,\frac12[$ for $\alpha$ to be in the range $]0,1[$, as assumed. 
But since $\Lambda=\mu/(1-n)$, the limiting case $n=1$ only makes sense if simultaneously $\mu = 0$, i.e. the total number of exogenous events 
grows sub-linearly with the sample length.

Another way to think about this limit, which may be more appropriate for financial applications, is to assume that $n = 1 -\delta$, 
$\delta \ll 1$. The autocorrelation function then decays as a power-law $\propto \tau^{-\alpha}$ up to a (long) crossover time $\tau^* \sim \delta^{-1/\epsilon}$,
beyond which it decays much faster, in order to make $\nu(\tau)$ integrable. 

What happens when $\epsilon > \frac12$ but $\varphi_0 = \epsilon$ such that $n=1$ (see Eq. (\ref{hawkes_power}))? Naively, the above analysis  predicts 
$\alpha < 0$, i.e. an autocorrelation that increases without bound as $\tau \to \infty$. This is of course impossible, and signals that in this case, the
critical Hawkes process is non-stationary. Such a process can be simulated by providing some past history with a constant average event rate 
$\Lambda$ for $t < 0$; its variance is then found to increase as $t^{2 \epsilon - 1}$ for $t > 0$.

The reason we have focused so much attention on this critical case is that, as we will show below, the empirical data clearly suggests that the relevant Hawkes 
process is indeed critical and stationary, i.e. precisely the Hawkes processes without ancestors considered by Br\'emaud \& Massouli\'e \cite{bremaud}. 


\begin{figure*}
\includegraphics[width=1\textwidth]{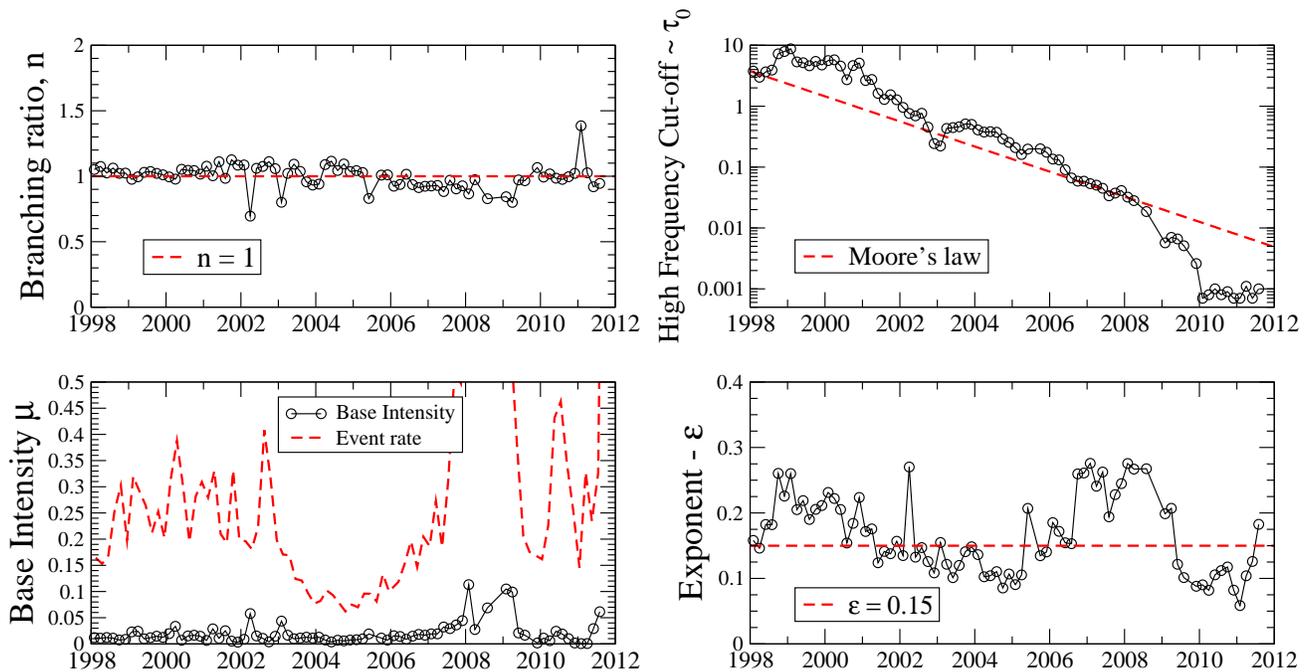}
\caption{\label{fig:mleresults} Results for the MLE fitting of the Hawkes process with power-law kernel to mid-point changes in the E-mini S\&P Futures contract. The fit is performed over two-month periods between Jan. 1998 and Dec. 2011. Top-Left: The branching ratio $n$ appears to change little and fluctuates around the critical value $n = 1$ for the 14-year period studied. Top-Right: The high frequency cut-off $\tau_0$, on the other hand, decreases exponentially over time as a result of the increasing frequency at which events are correlated, this trend is not too far out of line with Moore's law, the oft-cited observation that the processing speed of computers doubles every 18 months. Bottom-Left: The base intensity $\mu$ (in events per second) that has remained roughly constant over time, apart from spikes during crisis periods (for comparison, the average event rate is shown with the same scale). Bottom-Right: The exponent $\epsilon$ entering the power-law kernel $\phi(\tau)$, that has also remained roughly constant around $0.15$ -- see below.}
\end{figure*}

\section{Power-law versus Exponential Hawkes Kernel}
\label{sec:mle}

As mentioned above, FS estimated the parameters of a Hawkes processes on the midpoint changes of the E-mini S\&P Futures contract in the period 1998-2010 assuming an exponential kernel. However, this choice can only 
be justified if the dynamics of the market can be characterized by a single time scale, $\beta^{-1}$. This is at variance with many empirical studies that clearly elicit a scale-free, long-memory of measurements of market activity (such as trading volume and volatility). This strongly suggest the use of a power-law influence kernel. As a matter of fact, Bacry et al. \cite{bacry} have developed a non-parametric estimation technique that yields a power-law kernel over several decades for incoming market orders on both the Bund and the Dax  futures contracts. They find that the exponent $\epsilon$ (defined by Eq. (\ref{hawkes_power})) is small. 

We will revisit this non-parametric determination of the kernel in Sect. \ref{sec:nonparametric}, but for now we wish to repeat the analysis of \cite{filimonov} with the hypothesis of a simple power-law kernel, as given by 
Eq. (\ref{hawkes_power}). To allow for changes to the high frequency end of the Hawkes kernel over time resulting from improved temporal resolution of market events and reduced trading latency, we consider a power-law kernel with a smooth exponential cut-off for short lags. Furthermore, to expedite the calculation of the likelihood of the Hawkes process, we approximate the power-law as a sum of exponential functions with power-law weights \footnote{By approximating with a sum of exponentials we can take advantage of a recurrence relation which reduces the time complexity of the log-likelihood calculation in Eq. (\ref{eq:loglikelihood}) from $\mathcal{O}(N^2)$ to $\mathcal{O}(N)$}.
\begin{equation}
\phi(\tau | n, \epsilon, \tau_0) = \frac{n}{Z}\left(\sum_{i=0}^{M-1} \left(\frac{1}{\xi_i}\right)^{1+\epsilon} e^{-\frac{\tau}{\xi_i}} -S e^{-\frac{\tau}{\xi_{-1}}}\right)
\end{equation}
The scales of the exponentials are given by:
\begin{equation}
\xi_i = \tau_0 m^i \hspace{1pc} \textrm{for} \hspace{1pc} -1 \le i < M
\end{equation}
The parameter $Z$ is chosen such that $\int_0^\infty \phi(\tau) d\tau = n$ and $S$ such that $\phi(0) = 0$ (since we cannot expect market participants to react to events in zero seconds). We can vary the precision and extent of our power-law approximation by varying the parameters $m$ and $M$. For our calibration to the empirical data we have chosen $m=5$ and $M=15$, but we found our results to be robust with respect to variations in these parameters. For values of $\epsilon$ in the ranges we expect, the resulting function describes closely a power-law form with tail exponent $~\tau^{-(1+\epsilon)}$ but the negative exponential term provides a smooth drop to zero at lags shorter than approximately $\tau_0$.

Given observations of empirical market events (mid-point changes) at times $t_1,t_2, \ldots, t_n$ in an interval $[0,T]$ we can fit the model by maximising the log-likelihood \cite{rubin,ozaki} over the set of parameters $\theta = (\mu, n, \epsilon, b)$.
\begin{equation}
\log L(t_1,t_2,\ldots,t_n| \theta) = -\int_0^T \lambda(t| \theta) dt + \int_0^T \log \lambda(t|\theta) dN(t) \label{eq:loglikelihood}
\end{equation}

The data we use consists of a sequence of times in the period 1998-2011 when the mid-point of the bid and ask price of the E-mini S\&P contract was observed to change. Changes in this mid-point may have been the result of executed market orders or limit orders that have been placed or cancelled. Our data is time-stamped to the millisecond, but since sometimes multiple events occur during the same millisecond - the limit of our temporal resolution - we employ a randomisation procedure. We give each event a real number time-stamp distributed randomly and uniformly within the millisecond that it was reported. A similar operation is performed in \cite{filimonov}, but within the second resolution that they have. 

In fitting the model we only consider market events occurring during Regular Trading Hours (09:30 to 16:15). Furthermore, we take account of the `U-shape' intra-day activity profile. Without appropriate detrending, the fitting procedure would interpret the slow variation in event rate during the day as arising from the shape of the Hawkes kernel. To this end, we consider a detrended Hawkes process model
\begin{equation}
\lambda(t | \theta) = \frac{1}{w(t)}\left[\mu + \int_0^t w(s) \phi(t-s | \theta) dN(s)\right] \label{eq:detrendedintensity}
\end{equation}
where $w(t)$ is a periodic function defined over the trading day that dictates the weight to be appropriated to events occurring at different hours. The aim is to give less weight to events occurring during the morning and late afternoon when high activity is to be expected. The weights are chosen to be the reciprocal of the empirical average daily event rate at that moment of the day $w(t) = \frac{1}{R(t)}$ normalised such that $E[R(t)] = E[w(t)] = 1$.

The model parameters as a function of time are calculated between 1998 and 2011 on non overlapping two-month periods using an estimate of the function $R(t)$ made for each year. All regular trading hours in each two-month period are concatenated to form a single continuous point-process for which the log-likelihood of Eq. (\ref{eq:loglikelihood}) and Eq. (\ref{eq:detrendedintensity}) is maximised with respect to the four parameters $\theta = \{\mu, n,\epsilon, \tau_0\}$. The results are shown in Fig. \ref{fig:mleresults}.

Our key observation is that the branching ratio appears to change little over the course of the fourteen years studied but fluctuates about the critical value $n_c=1$ corresponding to the onset of stationarity. 
This means, that within the Hawkes framework, the majority of events observed in the market can be attributed to the long-range influence kernel, both now and in 1998. However, we see that the short lag cut-off of the kernel has decreased exponentially over time. This can be attributed to the steady emergence of higher frequency trading despite little change in the event rate. It may surprise the reader that the event rate has not increased significantly in the 14-year period (the event rate, for example in 2009 is comparable with that in 1999) despite the growth of the electronic futures market, but understand that the events studied are mid-point changes, not trades or order-book events and as such it is a better proxy for volatility than transaction quantity or volume.

Note that for the stationary critical Hawkes process with $n=1$ described in Sect. \ref{sec:critical}, the base intensity $\mu$ must be zero. The small ``excess'' base intensity observed in our results can be explained as arising from the contribution of the part of the power-law kernel that extends beyond the two-month window on which the fit is performed. The integral of this remainder is approximately $\epsilon \left(\frac{\tau_0}{T}\right)^\epsilon$ where $T \approx 2 \textrm{ months}$. Indeed, since each two month period is fitted independently, without prior event history, there must be some non-zero base intensity to explain the observation of any event at all!

\begin{figure}[h]
\includegraphics[width=1\columnwidth]{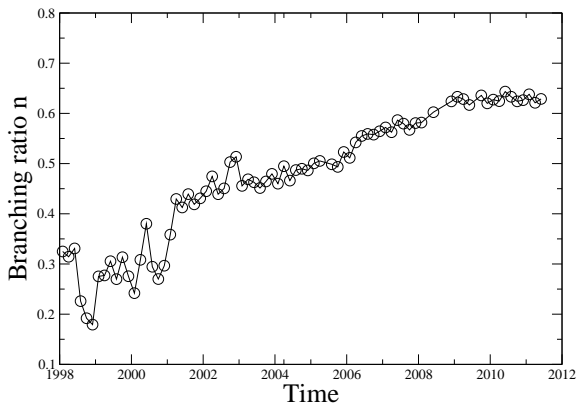}
\caption{\label{fig:recreation} The analysis of \cite{filimonov} on a critical power-law Hawkes processes. We generate synthetic power-law Hawkes processes with parameters $\{\mu,\epsilon,n\} = \{0.02,0.15,1.0\}$ but with a cut-off $\tau_0$ drawn from Fig. \ref{fig:mleresults} and attempt to fit these generated processes to a model with an exponential kernel on 30-minute windows. Furthermore we also impose the intra-second randomisation procedure employed in \cite{filimonov}. Despite the fact that the generated datasets represent critical Hawkes processes with $n = 1.0$, the analysis (incorrectly) reveals a sub-critical branching ratio which has been steadily increasing since 1998, as reported in \cite{filimonov}, Fig. 4-D.}
\end{figure}

Why is our result $n \approx 1$ over the whole period so different from the one presented in FS \cite{filimonov}? To reconcile our result with theirs, we propose that a sub-critical branching ratio $n < 1$ which increases over time can arise from the choice of an exponential short-memory model to fit a critical, power-law Hawkes process with a time dependent high-frequency cut-off $\tau_0$ on 30-minute windows. To this end, we have performed MLE fits of the exponential kernel model to synthetic data that we have generated using the power-law kernel hypothesis. We fixed $\{\mu, \epsilon, n\} = \{0.02, 0.15, 1.0\}$ (values which are approximately representative of the full fourteen years studied) but allowed the cut-off parameter $\tau_0$ to vary according to the MLE results of Fig. \ref{fig:mleresults}. For each two-month time period, we generated an ensemble of 100 power-law Hawkes processes from which, following FS's procedure, we drew a 30-minute window. After randomising the time-stamps within the second, as in \cite{filimonov}, the window was fit to the exponential kernel model in the same way as FS. The resulting branching ratio $\frac{\alpha}{\beta}$ was then averaged over all 100 windows in each ensemble. The results of this analysis are shown in Fig. \ref{fig:recreation}, and resemble much the data presented in FS (see their Fig. 4, panel D).

Despite the fact that the dataset is generated from a critical Hawkes process with $n = 1.0$, the analysis with the exponential kernel model on 30-minute windows reveals a sub-critical branching ratio which increases steadily over time. In 1998, since the market dynamics operated at a lower frequency, much of the contribution to $n$ from the integral of the power-law kernel lies outside the range of a 30-minute window, therefore the analysis with an exponential kernel on these fixed time windows only picks up the short-term ($<$ 30min) reflexivity which has increased in the past 14 years due to advancements in algorithmic and high frequency trading.

\section{Non-parametric estimation of the Hawkes Kernel}
\label{sec:nonparametric}

To complement and back up the results presented in Section \ref{sec:mle}, we also estimate the shape of the Hawkes kernel for several times in the 14-year period using non-parametric means. In \cite{bacry}, the authors present a methodology for estimating the kernel of multivariate Hawkes process based on Eq. (\ref{eq:covkernelrelation}) and the determination of the empirical auto-covariance of the event rate, Eq. (\ref{correlation}). We have repeated their analysis in the monovariate case for our data set, choosing $h=0.001s$ for 2009 (the maximum temporal resolution of our data), $h = 0.01s$ for 2005 and $h=0.1s$ for 1998. Following \cite{bacry} the auto-covariance is also discretely sampled at a rate $\Delta = h$ and the Fourier transform is performed in practice with DFT. The resulting kernel estimate is subsequently averaged with logarithmic bins to reduce noise for presentation.

\begin{figure}[h]
\includegraphics[width=1\columnwidth]{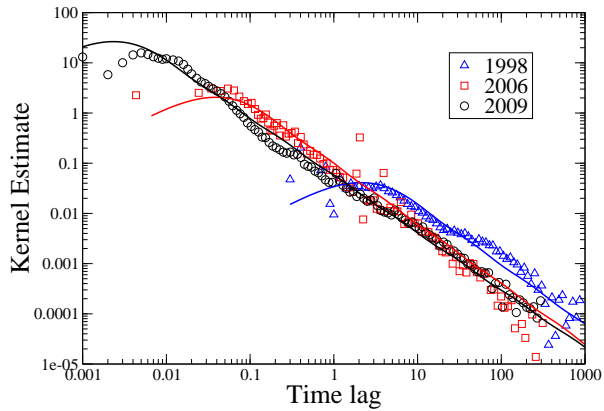}
\caption{\label{fig:nonparametric} Non-parametric kernel estimation of the Hawkes kernel in 1998, 2006 and 2009. The power-law form with decay exponent $\epsilon \approx 0.15$ is reproduced in all three years. However the high frequency cut-off of the kernels are considerably different. The kernel estimates match well with the MLE fits (solid lines) using the average parameter estimates for each year as taken from Fig. \ref{fig:mleresults}}
\end{figure}

To minimise the effects of intra-day seasonality, event rates are again detrended by dividing by the daily event rate as averaged for 5-minute bins over each year studied. Furthermore the empirical auto-covariance is estimated on 45-minute windows, during which we can expect the event rate to be relatively stationary anyway. The auto-covariance estimate used to produce the kernel in the range 1 ms -- 1000s in Fig. \ref{fig:nonparametric} is based on an average over all such 45-minute windows in a full year (1998, 2006 or 2009). 

The non-parametric procedure confirms the MLE results and also captures the change in the high-frequency cut-off between the two years. Note that the strong oscillations in the 1998 kernel estimate for $\tau < 1s$ are not random noise, but the result of a spurious correlation in how the data (although time-stamped to the millisecond) is collected or aggregated. However, the MLE estimation of the kernel shape seems robust to these data anomalies.

As an additional check of the goodness of fit of the power-law Hawkes process, we consider the residual process \cite{residuals}. We re-express the point process in terms of a transformed time $t \to t^*$ which is the integral of the conditional intensity.
\begin{equation}
 t^* = \int_{-\infty}^t \lambda(s)  ds \label{eq:transformedtime}
\end{equation}
where $\lambda(.)$ is calculated using the model on events prior to time $t$. If the Hawkes process with our chosen kernel, $\phi(\epsilon,n,\tau_0)$, is indeed a good description of the data, then the transformed process $N(t^*)$ should be Poisson with unit intensity. We assess this fit by analysis of the arrival times between events in the transformed time $t^*$, which if unit variance Poisson should be described by the law $P(\Delta t^*) = \exp{\left(-\Delta t^*\right)}$.

\begin{figure}[h]
\includegraphics[width=1\columnwidth]{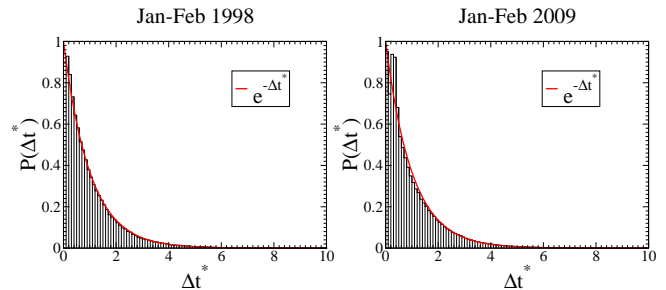}
\caption{\label{fig:residuals} The empirical distribution for the inter-arrival times of events in the transformed time of Eq. (\ref{eq:transformedtime}). 
The parameters for the power-law kernel used are extracted from the MLE results of Fig. \ref{fig:mleresults}
The Kolmogorov-Smirnov distance (i.e. the maximum deviation between the empirical CDF and the Poisson expectation) is 0.011 in 1998 and 0.031 in 2009.
}
\end{figure}

The results of this analysis for periods in 1998 and 2009 are given in Fig. \ref{fig:residuals}. 
The PDFs presented are produced from the inter-arrival times of the residual process for all points in the given two month periods. 
The kernel parameters used in the analysis are extracted from the MLE results of Fig. \ref{fig:mleresults}. 
These figures show that one can satisfactorily account for {\it all} events occurring in each two month period with a single exogenous intensity term 
and a three parameter kernel, which we consider in itself a remarkable fact. But even if the kernel shape and Hawkes dynamics were constant throughout 
each two month period, any deviation from the true kernel no matter how small would fail the Kolmogorov-Smirnov test due to the enormous number of points 
in our dataset. 

\section{The shape of the Hawkes Kernel at large lags}

The results in the previous section demonstrate that a Hawkes process described by a power-law kernel with an exponent $\epsilon \approx 0.15$ is indeed a good fit to the data in the millisecond to 5-minute region. We will now show that there is a marked regime change in the nature of the correlations for larger time horizons. To see this, we use Detrended Fluctuation Analysis (DFA) \cite{dfa}, a commonly used tool for detecting long-range correlations in the presence of trends and non-stationarity. More precisely, we compute:
\begin{equation}
 F(L) = \sqrt{\frac{1}{T} \sum_{t=1}^T \left(N(t) - f_L(t)\right)^2}
\end{equation}
where $N(t)$ is the empirical event counting process formed by concatenating all regular trading hour periods in 1998-2011. $N(t)$ is discretised into $T$ 0.1 second bins and $f_L(t)$ is a piecewise function formed by the linear regression to the points $N(t)$ in non-overlapping windows of length $L$ which cover the period studied. If $N(t)$ behaved as a Brownian motion with uncorrelated fluctuations, as for a simple Poisson process, we would expect $F(L) \sim L^H$ with a Hurst exponent $H = \frac12$. In the presence of long-ranged correlations induced by a power-law Hawkes kernel, one finds: 
\begin{equation}
H = \frac12 + \epsilon, \qquad (0 < \epsilon < \frac12) \label{eq:counterintuitive}
\end{equation}
Note that this result is not intuitive, since the larger the value of $\epsilon$, the faster the decay of the Hawkes kernel, but the stronger the deviation from the Poisson result $H=\frac12$. This paradoxical behaviour 
was in fact already apparent in the relation $\alpha = 1 - 2\epsilon$ derived in Sect. \ref{sec:critical} above.

\begin{figure}[h]
\includegraphics[width=1\columnwidth]{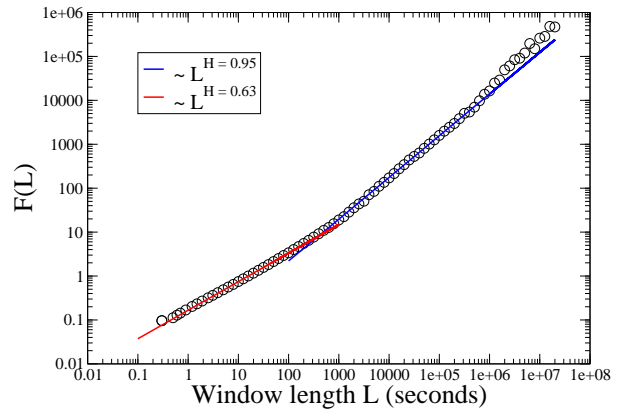}
\caption{\label{fig:dfa} Detrended Fluctuation Analysis on the counting process for market events in the period 1998-2011. We identify two regimes, one at high frequencies corresponding to a Hurst exponent of $H \approx 0.63$ and one at much lower frequencies corresponding to $H \approx 0.95$.}
\end{figure}

Fig. \ref{fig:dfa} clearly shows two regions. For small time windows ($L < 1000 s$), the Hurst exponent is found to be $H \approx 0.63$, consistent with the observation of $\epsilon \approx 0.15$ in the previous section. For longer time windows ($L > 1000 s$), the diffusion of the integrated event rate becomes more super-diffusive and leads to a stronger Hurst exponent of $H \approx 0.95$. Such a change to a more super-diffusive regime has been observed before for measures of market volatility \cite{liu,fengzhong}. However these previous studies have reported the cross-over at the scale of one day, whereas in our dataset for event rate it is happening much faster (10 - 20 minutes). We have repeated the same analysis for absolute value of five-minute returns and confirmed that the cross-over time is indeed longer in that case, on the scale of a day.

We attribute the difference to the fact that the distribution of five-minute returns is well known to have fat-tails, much fatter than the distribution of the number of price changes over five minutes. The extra noise brought about by these tails reduce short-time correlations, and therefore bring the value of $H$ closer to $1/2$. 

\begin{figure}[h]
\includegraphics[width=1\columnwidth]{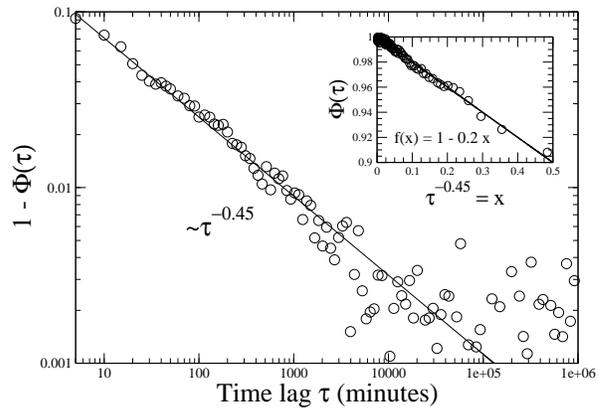}
\caption{\label{fig:5minkernel} The result of non-parametric kernel estimation on the auto-covariance of event rate as calculated using 5-minute bins for the full 14-year period. We plot $1 - \Phi(\tau) = 1 - \int_0^{\tau} \phi(s) ds$ in a log-log representation in the main figure and $\Phi(\tau)$ as a function of $\tau^{-0.45}$ in the inset, clearly demonstrating that $\phi(s)$ integrates to unity and has a power-law tail. This tail exponent $\epsilon=0.45$ is consistent with the DFA analysis in Fig. \ref{fig:dfa}.}
\end{figure}

The critical power-law Hawkes kernel implied by the stronger autocorrelation $H \approx 0.95$ corresponds to a larger decay exponent of $\epsilon \approx 0.45$. 
To visualise the shape of the Hawkes kernel at these large time scales, we estimate the auto-covariance of event rate but now using 5-minute binned data over the full 14-year period. To do this, we again treat the event rate dataset as one large continuous time series by concatenating all regular trading periods. The resulting kernel estimate is shown in Fig. \ref{fig:5minkernel}. Two results can be read from this figure: a) again, the Hawkes kernel indeed integrates
to unity, confirming the critical nature of the process implied by the long-memory of event rates (as discussed in Sect. \ref{sec:Hawkes}); b) the exponent $\epsilon=0.45$ extracted from the DFA analysis correctly describes the time dependence 
of the Hawkes kernel in the region $\sim 1000$ seconds $\to$ $10^6$ seconds (40 days). The value $\epsilon=0.45$ leads to $\alpha \approx 0.1$, which is indeed similar to the value of the exponent for the decay of volatility correlations found in the literature (see e.g. \cite{MbW,Chicheportiche}). 

\section{Conclusion}

In this study, we have confirmed that Hawkes processes provide a useful framework to describe the high-frequency activity in financial markets. However, contrary to Filimonov \& Sornette \cite{filimonov}, but in agreement with Bacry et al. \cite{bacry}, we have found that the Hawkes kernel (that describes how past events influence future activity) decays as a power-law, rather than exponentially. The branching ratio we extract from the calibration of our model to the mid-point changes in the E-mini S\&P over 14 years is close to critical and has remained constant over time, 
contrary to the claim of Filimonov \& Sornette that reflexivity has substantially increased since 1998. Nonetheless, we uncover that the short lag cut-off of the power-law kernel which we fit has 
significantly decreased over time, with the emergence of higher frequency trading and higher temporal resolution in the market. We have argued that this explains the discrepancy 
between our results and those of Filimonov \& Sornette. Using both a non-parametric method and Detrended Fluctuation Analysis, we have found that the Hawkes kernel must in fact be described by {\it two} power-laws: one decaying as $\approx \tau^{-1.15}$ for 
short time scales ($\tau < 1000$ seconds), crossing over to a faster decay $\approx \tau^{-1.45}$ for longer time scales. This long time regime is consistent with the slow decay of volatility correlations reported in the literature.

The picture which emerges from our analysis is one in which bursts of diverging trading activity are just as inevitable now as they were in 1998 (and probably as far back as financial markets existed), but that the time-scale over which they may occur has substantially shortened in the last decade. We have shown that the Hawkes kernel must indeed be critical whenever the event arrival is a long-memory process. Based on this observation and on the results of Bacry et al. \cite{bacry} on the Bund and the DAX futures, 
we conjecture that our critical market scenario is common to most actively traded markets. Interestingly, financial markets appear to be well described by a critical Hawkes process ``without ancestors'', a theoretical scenario put forth by Br\'emaud \& Massouli\'e \cite{bremaud}. In non-technical terms, this means that the rate of endogenous events in financial markets is overwhelming compared to that of exogenous (news related) events. A detailed mechanism explaining why markets are precisely poised at criticality is however still lacking.

\acknowledgements
We thank Marc Potters, Julianus Kockelkoren, Emeric Henry, Yves Lemp\'eri\`ere, Joachim de Lataillade, Cyril Deremble and Iacopo Mastromatteo for many helpful discussions as well as Emmanuel Bacry and Jean-Fran\c{c}ois Muzy.

\bibliography{hawkespaper}

\end{document}